# Dynamic channel selection in wireless communications via a multi-armed bandit algorithm using laser chaos time series

Shungo Takeuchi[1*], Mikio Hasegawa[1], Kazutaka Kanno[2], Atsushi Uchida[2], Nicolas Chauvet[3], and Makoto Naruse[3**]

1 Department of Electrical Engineering, Tokyo University of Science, 6-3-1 Niijuku, Katsushika-ku, Tokyo 125-8585, Japan

2 Department of Information and Computer Sciences, Saitama University, 255 Shimo-Okubo, Sakura-ku, Saitama City, Saitama 338-8570, Japan

3 Department of Information Physics and Computing, Graduate School of Information Science and Technology, 7-3-1 Hongo, Bunkyo-ku, Tokyo 113-8656, Japan

Email: * s-takeuchi@haselab.ee.kagu.tus.ac.jp, ** makoto_naruse@ipc.i.u-tokyo.ac.jp

**Abstract**

Dynamic channel selection is among the most important wireless communication elements in dynamically changing electromagnetic environments wherein, a user can experience improved communication quality by choosing a better channel. Multi-armed bandit (MAB) algorithms are a promising approach that resolve the trade-off between channel exploration and exploitation of enhanced communication quality. Ultrafast solution of MAB problems has been demonstrated by utilizing chaotically oscillating time series generated by semiconductor lasers. In this study, we experimentally demonstrate a MAB algorithm incorporating laser chaos time series in a wireless local area network (WLAN). Autonomous and adaptive dynamic channel selection is successfully demonstrated in an IEEE802.11a-based, four-channel WLAN. Although the laser chaos time series is arranged prior to the WLAN experiments, the results confirm the usefulness of ultrafast chaotic sequences for real wireless applications. In addition, we numerically examine the underlying adaptation mechanism of the significantly simplified MAB algorithm implemented in the present study compared with the previously reported chaos-based decision makers. This study provides a first step toward the application of ultrafast chaotic lasers for future high-performance wireless communication networks.

# Introduction

The resources for wireless communications are physically limited because of their narrow frequency bandwidth and ever-increasing demands in society [1]. Therefore, dynamic channel selection is among the most important wireless communication elements in dynamically changing electromagnetic environments such that a user can experience improved communication quality by choosing a better channel in terms of, for instance, the communication throughput [2]. The metric could also minimize energy consumption, communication delay, etc., depending on the interest of a given system. Further, autonomous and prompt adaptation is important for dynamically changing wireless communication environments.

Lai *et al.* modelled the channel selection problem as a multi-armed bandit (MAB) problem [3]; an example of a MAB problem is finding the most highly profitable slot machine among many machines. To find the best machine, one must conduct a search for a high reward machine. However, too much exploration may accompany significant losses whereas, a too quick decision may result in missing the best choice. Hence, a difficult trade-off exists, which is referred to as an exploration–exploitation dilemma [4]. A channel selection problem in wireless networks can be regarded as a MAB problem by associating the communication quality (such as the throughput) to the reward of a slot machine. Recently, Kuroda *et al.* applied a Tug-of-War algorithm [5] for MAB problems to a wireless local area network (WLAN) to demonstrate its effectiveness [6]. In a wider context, Obayuiwana *et al.* reviewed network selection problems using a decision-making algorithm [7].

Meanwhile, ultrafast solution of MAB problems has been demonstrated by utilizing chaotically oscillating waveforms generated by semiconductor lasers [8]. The negative autocorrelation inherent in laser chaos yielded an accelerated solving of a two-armed bandit problem, achieving 1 ns latency. Furthermore, by employing a time-domain multiplexing technique, scalable decision-making has been demonstrated up to a 64-armed bandit problem [9]. Such physical approaches exploit unique ultrafast physical attributes of laser physics aiming at deriving ultimate decision-making performance beyond the operational limit of conventional signal processors.

While current photonic decision-making systems still require high-end experimental facilities, photonic and optoelectronic devices as well as integration technologies are rapidly progressing [10]. Homma *et al.* recently demonstrated an on-chip photonic decision maker using a ring laser [11]. Therefore, examining the applicability and feasibility of such ultrafast photon-based decision-making technology to wireless communication is an interesting and indispensable step toward future innovative developments. In addition, the experimental results shown by application studies are useful from a photonic technology perspective to understand critical specifications toward future research.

From these standpoints, this study experimentally implemented a MAB algorithm incorporating laser chaos time series in a WLAN environment. An autonomous and adaptive dynamic channel selection was successfully demonstrated based on an IEEE802.11a-based, four-channel WLAN system. While the chaotic laser time series was arranged prior to the WLAN experiments, the results confirm the usefulness of ultrafast sequences for real wireless applications. In addition, we numerically examined the underlying adaptation mechanism of the significantly simplified MAB algorithm employed in the present study, compared to previously reported chaos-based decision makers.

**Results**
**Principle**
We examined the following scenario to assess the laser-chaos-based MAB algorithm for dynamic channel selection. Assume a wireless communication access point (AP), which is a WLAN system. The objective is to realize the best channel selection for an edge device (or a terminal) such that the throughput is maximized wherein, the best channel may change over time because of changing network traffic as schematically illustrated in Fig. 1. The throughput is periodically evaluated to measure the channel quality. In the MAB problem formation, a positive reward is given when the chosen channel provides an improved throughput compared to the average throughput over time. Otherwise, a negative reward is provided (details are discussed in the following and the *Methods* section.) The interest of the present study is to implement a MAB algorithm incorporating laser chaos time series and confirm its operational ability.

It should be emphasized that while the channel selection strategy examined in the present study is based on the scalable decision-making methods using chaotic time series demonstrated in Ref. [9], a large part of the method has been updated to accommodate the real wireless application; these methods are summarized in the following three items. The first is that here, by experimentally introducing external traffic, we do consider a dynamically changing environment, whereas Ref. [9] does not. The second is that the notion of reward is transformed to the communication throughput; a reward is dispensed at '*t*' if the resulting throughput is larger than the average throughput over time. The third is, stemming from the above two points, the MAB method has been significantly simplified compared with the one presented in Ref. [9]. Further, the underlying mechanism of the method is examined in detail.

**Chaos-based decision-making**
Here, we introduce the chaos-based decision making in the channel selection. A semiconductor laser typically provides a stable intensity light profile, but if it is subject to optical feedback to its cavity via an external time delay, referred to as delayed feedback, the light oscillation destabilizes and shows a chaotically oscillating waveform, which has been comprehensively studied in the literature [12, 13] (Fig. 2a). The photonic dynamics provide an ultrafast irregular time sequence unachievable by electronics means; hence, application to photonic ultrafast random number generation [14, 15] and photonic reservoir computing [16] has been demonstrated. Naruse *et al.* exploits such photonic physical processes in decision-making [8, 9, 17].

The principle of the chaos-based decision-making is as follows. Let the intensity level of the laser chaos sequence at time *t* be *s*(*t*). It is then compared to a threshold level, $TH_1$ as shown in Fig. 2a. If the intensity level *s*(*t*) is less than or equal to $TH_1$, the decision is to choose Option 0 (otherwise Option 1). If a reward is dispensed by selecting Option 0, the threshold $TH_1$ is slightly increased such that the likelihood of choosing Option 0 increases in the subsequent trial. It is critical to check the alternative selection; this is realized by the chaotic dynamics of the incoming signal. The negative autocorrelation yields enhanced performance in finding a better decision [8]. To extend this to MAB problems greater than three arms, the aforementioned principle is cascaded in a pipeline manner as shown in Fig. 2b; the first sampling regards the most significant bit (MSB) of the selections, and the second decides the second MSB [9]. Chaotic time series have been experimentally solved up to a 64-armed bandit problem [9].

In such decision-making mechanisms, the estimated reward probability $\hat{P}_i$ of the selection *i*, has been considered important in former studies [5, 8, 9]. The increase or decrease in the threshold level should be attenuated based on $\hat{P}_i$. In the case of a two-armed bandit problem, for instance, if the chosen decision yields a reward, the threshold level is increased (or decreased) in unit 1, otherwise the threshold is decreased (or increased) by the following amount

$$\Omega = \frac{\hat{P}_0 + \hat{P}_1}{2 - (\hat{P}_0 + \hat{P}_1)} \tag{1}$$

to accurately converge to the best decision making [8]. The threshold level is then updated using the following rule:

$$TH_1 = \begin{cases} \alpha \times TH_1 + 1 & \text{if reward is dispensed} \\ \alpha \times TH_1 - \Omega & \text{if reward is not dispensed} \end{cases}, \tag{2}$$

when the decision is to select the Option 0. It is remarkable that the amount of change in the threshold is different when the reward is dispensed and when it is not, as shown in Eq. (2). The theoretical detail can be found in [18], but an intuitive picture is the following. Assume that the reward probabilities of the two slot machines are 0.3 and 0.2. In such a case, the probability of *winning* is only (0.3 + 0.2) / 2 = 1/4, whereas that of *losing* is 3/4. That is to say, the threshold update, in the case of losing, should be three times more attenuated, otherwise the decision does not converge to a correct one. Here, the parameter $\alpha$ is termed the forgetting parameter to adapt to the uncertain environmental change [8, 9]; a smaller $\alpha$ indicates rapid forgetting of past threshold values. In the case of four-armed bandits demonstrated in the following, a total of five threshold values are arranged.

The laser chaos sequences were sampled using a high-speed digital oscilloscope at a rate of 100 G sample/s (at a 10 ps sampling interval) with 10,000,000 points at an 8-bit resolution. These 10-M data points were stored 120 times for each signal train; hence, 120 types of 10-M-long sequences were stored. The detailed experimental conditions are described in the *Methods* section.

**System implementation**

We implemented the aforementioned chaos-based MAB algorithm in an edge device (or a terminal) that communicates with an access point (AP) based on the IEEE802.11a protocol of the WLAN. In the system, we prepared four available channels specified by the channel identities of 36, 40, 44, and 48 in IEEE802.11a and corresponding to the frequency bands of 5.18, 5.20, 5.22, and 5.24 GHz, respectively (Fig. 2b). As mentioned in the *Introduction*, the laser chaos sequences were obtained prior to the WLAN experiments; the chaotic sequences are stored in the memory of the terminal device, from which the time sequences are sequentially sampled to decide the channel to select. Furthermore, to clearly examine the ability for autonomous channel selection, we applied external traffic by preparing another terminal device to deteriorate the throughput, or induce intentional traffic congestion in the corresponding channels. The details of the implementation are described in the *Methods* section.

**Dynamic channel selection**

By configuring the external traffic-generating terminal, the best channel, namely, the most vacant or high-throughput channel, is periodically changed in the order of channel 48, 44, 40, and 36 at an interval of approximately 40 s, which corresponds to 50 cycles in terms of the terminal channel updating cycle. Figure 3a shows the time evolution of the channels chosen by the terminal driven by the chaos-based MAB algorithm. While the terminal is selecting the non-best channels particularly during the durations immediately after the external traffic environment is changed (that is, the initial moments and the periods shortly after the 50, 100, and 150 cycles), the terminal selects channel 48 during the first 50 cycles, followed by channels 44, 40, and 36 during the subsequent cycles. That is, dynamic channel selections were successfully demonstrated. Likewise, Fig. 3b summarizes the throughput time evolution, more precisely, the Transmission Control Protocol (TCP) throughput, between the terminal and AP.

The throughputs are degraded during the initial moments and several cycles after 50, 100, and 150 cycles because of the uncertainty in the terminal knowing the situation of the wireless environment. The throughput reaches greater than 10 Mbps, demonstrating a higher throughput in a dynamically changing electromagnetic environment.

**Analysis of the simplified algorithm**

During the experimental demonstration, we did *not* estimate the reward probability of the options. Instead, we used a fixed value of $\Omega$ equal to unity regarding Eqs. (1) and (2). Meanwhile, in the present study, we assigned a significantly small value for the forgetting parameter $\alpha = 0.9$ unlike larger values such as $\alpha = 0.99$ employed in previous studies [9]. This is an interesting and important aspect because a greatly simplified algorithm reduces the computational latency and costs including energy consumption for computation. We numerically examined the underlying mechanism as to why non-estimates of the probability provide successful dynamic channel selections through the following two-armed analysis.

Let the reward probabilities of two slot machines, $P_0$ and $P_1$, interchange with each other every 2,500 cycles. A total of 10,000 cycles of decision-making are consecutively executed. The 10,000 cycles are repeated 12,000 times. Then, we evaluate the correct selection rate (CSR), which is the ratio of selecting the higher-reward-probability choices versus the number of repetitions. The average CSR throughout the 10,000 cycles was examined. We investigated the following three problems:

[Problem 1] Reward probabilities are given by $(P_0, P_1) = (0.1, 0.9)$ or vice versa.

[Problem 2] Reward probabilities are given by $(P_0, P_1) = (0.5, 0.9)$ or vice versa.

[Problem 3] Reward probabilities are given by $(P_0, P_1) = (0.1, 0.2)$ or vice versa.

Figure 4 summarizes the average CSRs for Problems 1–3 with and without reward probability estimations, which correspond to the *fixed* $\Omega$ and *flexible* $\Omega$ strategies, respectively. The forgetting parameter is configured as $\alpha = 0.9$ and $\alpha = 0.99$ as shown in Fig. 4a and 4b, respectively.

In Problem 1, notably, the accurate $\Omega$ is 1 because $P_0 + P_1 = 1$ in Eq. (1), meaning that the fixed $\Omega = 1$ actually reflects the accurate probability estimate. Therefore, the CSR average via the fixed $\Omega$ strategy shows a large value of nearly unity as shown in Fig. 4a(1) and 4b(1). In fact, as the flexible $\Omega$ strategy accompanies certain exploration processes, the fixed $\Omega$ strategy outperforms the flexible strategy.

In Problem 2, the fixed $\Omega$ strategy does not utilize the correct $\Omega$ (which is approximately 2.33); however, the resultant CSR average nearly shows unity when the forgetting parameter is small ($\alpha = 0.9$) (Fig. 4a(2)). Even with a large forgetting parameter ($\alpha = 0.99$) (Fig. 4b(2)), the fixed approach is superior to the flexible approach; this is presumably because the decision-making problem is not too difficult, the difference in the probability is 0.4, and hence, a fixed $\Omega$ with rapid forgetting does work. Indeed, in Problem 3, wherein the reward probability difference is only 0.1, the fixed $\Omega$ method does *not* provide good CSR values either with a smaller or a larger forgetting parameter as shown in Fig. 4a(3) and 4b(3), respectively. From these analyses, the fixed $\Omega$ does work when the forgetting parameter is configured as a small value as well as when the differences between the best selection and the next is minute.

**Conclusion**

In this study, dynamic wireless channel selection was demonstrated based on a MAB algorithm operated with experimentally obtained laser chaos time sequences. We implemented the proposed system in an IEEE802.11a-based four-channel WLAN and successfully demonstrated autonomous and adaptive channel selection to obtain higher throughputs greater than 10 Mbps between the terminal and access points in dynamically changing network traffic. In addition, we have performed dynamic channel selection through significantly simplified methods that reduced computational costs by eliminating the estimation of reward probabilities of the environment. We numerically examined its underlying mechanism showing that rapid forgetting of past parameters provides an even better performance when the given channel selection situation is not too difficult. Though the present study utilizes laser-chaos-based decision making in a terminal device for the purpose of maximizing throughputs, other interesting scenarios can be considered ranging from network selections, allocations of frequencies, or computational resources in wireless networks, data centres, etc. This study provides a first step toward compositionality of ultrafast photonic systems and wireless communication applications.

**Methods**

**Laser chaos.** A semiconductor laser (NTT Electronics, KELD1C5GAAA) operated at a centre wavelength of 1547.785 nm was coupled with a polarization-maintaining coupler. The light was connected to a variable fibre reflector, providing delayed optical feedback to the laser. A chaotic oscillatory light intensity profile was generated by setting the variable reflector to feedback 80 µW of optical power to the laser. The fibre length between the laser and reflector was 4.55 m (with a delay time of 43.8 ns). The output light was detected using a high-speed, alternating-current-coupled photodetector (New Focus, 1474-A) through an optical isolator and optical attenuator, which was sampled using a digital oscilloscope (Tektronics, DPO73304D) at 100 G sample/s.

**Chaos-based decision making and dynamic channel selection.** The four wireless communication channel selections are denoted by binary digits as follows: $D_1D_2 = \{00, 01, 10,$ and $11\}$. In chaos-based decision-making, the first sampling from the chaotic time sequence determines the MSB of the selections ($D_1$), and the next provides the second MSB ($D_2$). (The general formula is provided in ref. [9].) The chaotic signal $s(t_1)$ measured at $t = t_1$ is compared to a threshold value $TH_1$. If $s(t_1)$ is less than or equal to the threshold $TH_1$, the MSB decision is 0, denoted by $D_1 = 0$. Otherwise, the MSB is 1 ($D_1 = 1$). The chaos signal level $s(t_2)$ measured at $t = t_2$ is subjected to another threshold value denoted by $TH_{2,0}$ when $D_1 = 0$. If $s(t_2)$ is less than or equal to the threshold value $TH_{2,0}$, the second MSB is 0 ($D_2 = 0$). Otherwise, the second MSB is 1 ($D_2 = 0$). Finally, the decision to select is determined by $D_1D_2$.

Checking the TCP throughput of the chosen channel, if it is greater than the average TCP throughput over time, we consider that the decision is correct or a reward is dispensed. Accordingly, the MSB threshold value is updated as follows:

$$\begin{aligned} TH_1(t+1) &= \alpha TH_1(t) + 1 \quad \text{if } D_1 = 0 \\ TH_1(t+1) &= \alpha TH_1(t) - 1 \quad \text{if } D_1 = 1 \end{aligned} \tag{3}$$

while the threshold for the second MSB is revised as follows:

$$\begin{aligned} TH_{2,0}(t+1) &= \alpha TH_{2,0}(t) + 1 \quad \text{if } D_1 = 0, D_2 = 0 \\ TH_{2,0}(t+1) &= \alpha TH_{2,0}(t) - 1 \quad \text{if } D_1 = 0, D_2 = 1 \end{aligned} \tag{4}$$

or

$$\begin{aligned} TH_{2,1}(t+1) &= \alpha TH_{2,1}(t) + 1 \quad \text{if } D_1 = 1, D_2 = 0 \\ TH_{2,1}(t+1) &= \alpha TH_{2,1}(t) - 1 \quad \text{if } D_1 = 1, D_2 = 1 \end{aligned} \tag{5}$$

depending on the MSB value.

If the instantaneous TCP throughput becomes smaller than the TCP throughput averaged over time, we consider that the decision is wrong. Accordingly, the threshold values are updated in the opposite directions compared to those of Eqs. (3) to (5). That is, the MSB threshold is updated as follows:

$$\begin{aligned} TH_1(t+1) &= \alpha TH_1(t) - 1 \quad \text{if } D_1 = 0 \\ TH_1(t+1) &= \alpha TH_1(t) + 1 \quad \text{if } D_1 = 1 \end{aligned} \tag{6}$$

while the second MSB threshold is revised as follows:

$$\begin{aligned} TH_{2,0}(t+1) &= \alpha TH_{2,0}(t) - 1 \quad \text{if } D_1 = 0, D_2 = 0 \\ TH_{2,0}(t+1) &= \alpha TH_{2,0}(t) + 1 \quad \text{if } D_1 = 0, D_2 = 1 \end{aligned} \tag{7}$$

or

$$\begin{aligned} TH_{2,1}(t+1) &= \alpha TH_{2,1}(t) - 1 \quad \text{if } D_1 = 1, D_2 = 0 \\ TH_{2,1}(t+1) &= \alpha TH_{2,1}(t) + 1 \quad \text{if } D_1 = 1, D_2 = 1 \end{aligned}. \tag{8}$$

As mentioned in Eq. (2), the number of changes in the threshold can be configured differently as specified by $\Omega$, reflecting the estimated reward probabilities of the selections [8, 9]. In the present study; however, a fixed number of changes is employed while introducing a smaller forgetting parameter $\alpha = 0.9$. The actual threshold parameter values are defined by the nearest integers of ($TH_1$, $TH_{2,0}$, $TH_{2,1}$) multiplied by a constant value so that they can take five levels. The details of the flexible updating of $\Omega$ and the thresholds are found in Ref. [9].

**Experimental wireless communication implementations.** The four-channel WLAN

system was constructed as follows. The AP was implemented by combining four wireless communication terminals (ELECOM, WDC-433DU2HBK) to a microprocessor unit (Raspberry PI 3 Model B). The edge device, or terminal, where the laser-chaos-based MAB algorithm was installed was a personal computer (HP ProBook 4340s, 1.9-GHz central processing unit (CPU), 4-GB random-access memory (RAM)) with an Ubuntu operating system (OS). The MAB algorithm was implemented using a shell script while the throughput measurement was realized using the command iperf. Intentional load traffic was generated by three other personal computers (which were also ProBook 4340s models) connected to separately arranged APs (Buffalo, AirStation WCR-1166DS) nearby.

**Numerical studies regarding the simplified algorithm.** The simulation studies regarding the simplified algorithm employed in the experiments and the former algorithm with reward probability estimation were conducted using a personal computer (Lenovo H520s, 47466AJ, 3-GHz CPU, 4-GB RAM, Windows 10 Professional OS). MATLAB 2017a was used for coding.

**Data availability.** The datasets generated during the current study are available from the corresponding author on reasonable request.

**Acknowledgements**

This work was supported in part by the CREST project (JPMJCR17N2) funded by the Japan Science and Technology Agency, the Core-to-Core Program A. Advanced Research Networks and Grants-in-Aid for Scientific Research (JP17H01277 and JP19H00868) funded by the Japan Society for the Promotion of Science.

**Author Contributions**

M.N. and M.H. directed the project. K.K. and A.U. conducted the laser chaos experiments. S.T., M.N. and M.H. designed the wireless communication systems. S.T. and M.N. performed the signal processing. S.T., M.N., K.K., A.U., N.C. and M.H. analysed the data. S.T. and M.N. wrote the paper.

**Additional Information**

**Competing interests:** The authors declare no competing interests.

**Correspondence and requests for materials** should be addressed to S.T or M.N.

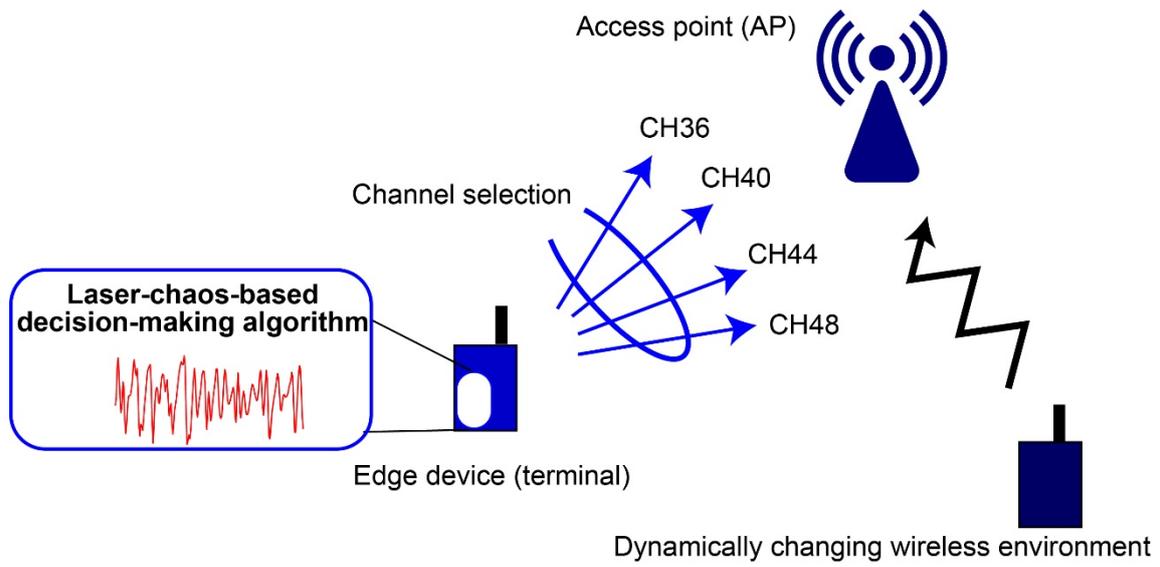

**Figure 1.** Dynamic channel selection using the laser-chaos-based decision-making algorithm in a dynamically changing wireless communication environment.

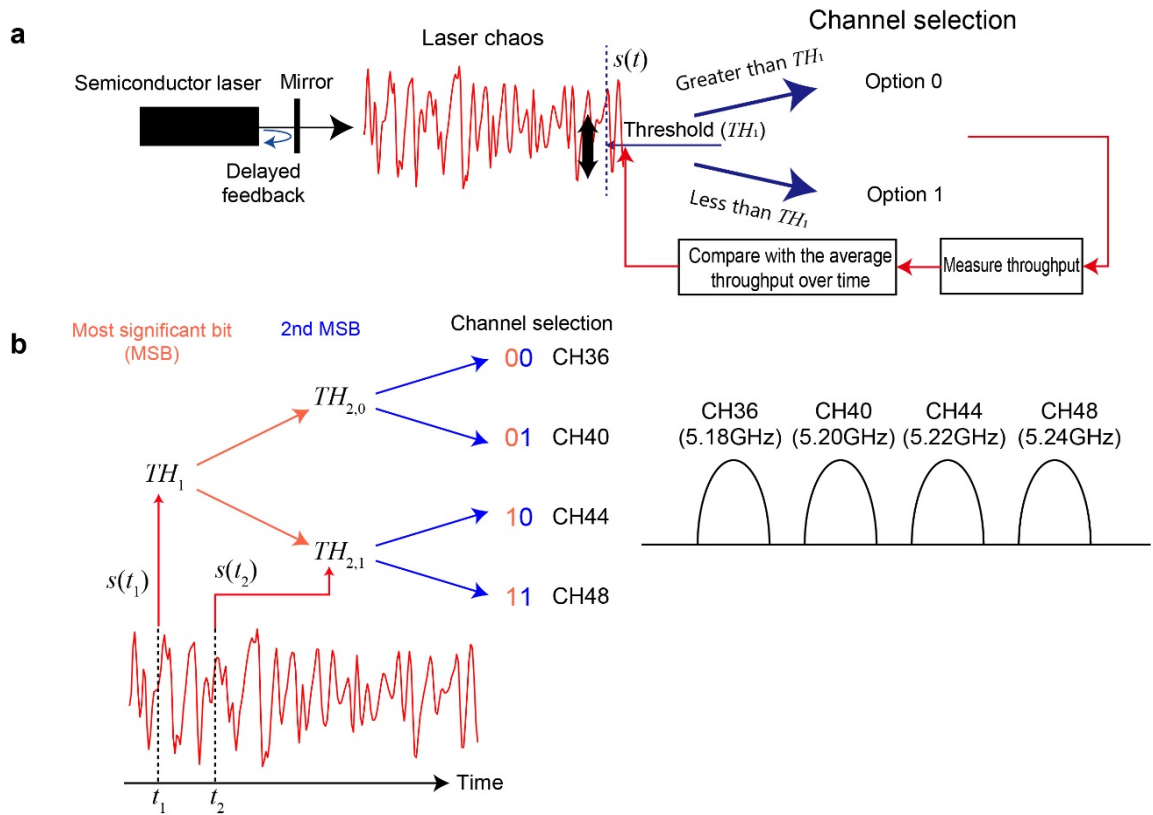

**Figure 2.** Channel selection principle using chaotic time series generated by semiconductor lasers. (**a**) Comparison between the signal level of the chaotic sequence and a dynamically reconfigured threshold level provided channel selection (either Option 0 or Option 1). If the throughput of the chosen channel is larger than the average throughput over time, a reward is dispensed. (**b**) Channel selection among multiple choices is performed via time-domain multiplexing. The present work studied the dynamic selection from four channels in IEEE802.11a.

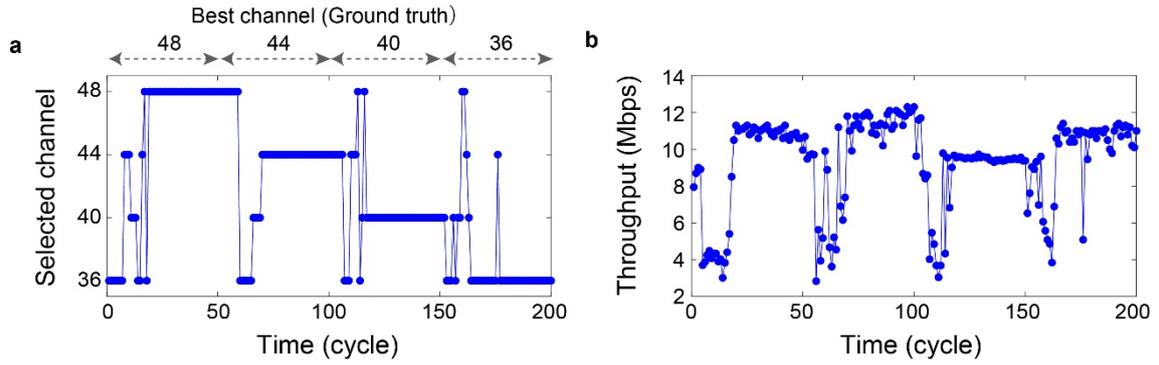

**Figure 3.** Experimental demonstration of dynamic channel selection utilizing chaotic laser sequences. The best channel with the highest throughput is configured as 48 during the first 50 cycles followed by 44, 40, and 36 during the subsequent 50 cycles, respectively. (**a**) Time evolution of the selected channel via the edge device supported by the laser-chaos-based MAB algorithm. Autonomous adaptation is successfully demonstrated. (**b**) Time evolution of the observed TCP throughput. Greater than 10-Mbps throughputs were accomplished throughout the operation except during the initial cycles and the cycles after the environmental changes at 50, 100, and 150 cycles.

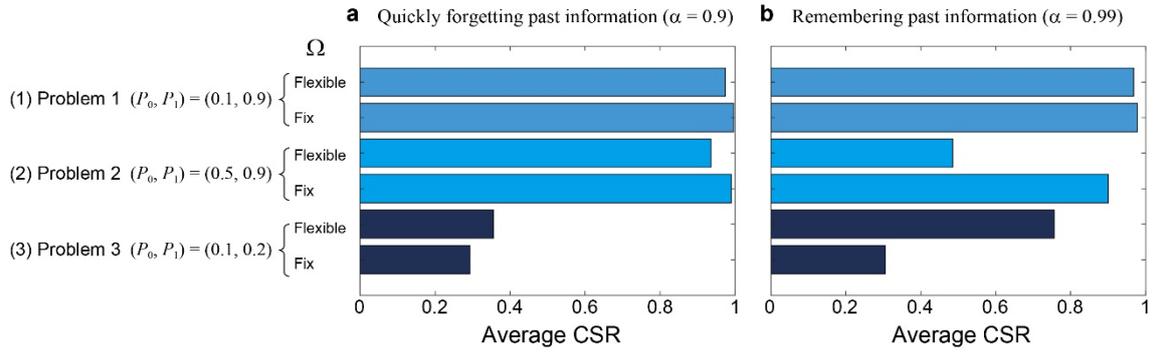

**Figure 4.** Analysis of CSR using the fixed Ω parameter in the MAB algorithm. (**a**) By quickly forgetting past records, correct decision making was conducted in Problems 1 and 2. (**b**) With regard to a difficult decision problem defined by Problem 3, the fixed Ω parameter approach yielded poor performance.